\documentclass[twoside,a4paper,11pt]{sea22}
\usepackage{graphicx}
\usepackage{hyperref}
\usepackage{movie15}
\usepackage{color}
\usepackage{siunitx}  
\usepackage{natbib}    
\usepackage[export]{adjustbox}
\def\hoy{\number\day \space de \space\ifcase\month\or
 Enero\or Febrero\or Marzo\or Abril\or Mayo\or Junio\or
 Julio\or Agosto\or Septiembre\or Octubre\or Noviembre\or Diciembre\fi
 \space de \number\year}
\def\ii/{\'{\i}}
\def\cion/{ci\'on}
\def\cao/{\c c\~ao}
\def\utw{\smash{\rlap{\lower5pt\hbox{$\sim$}}}}
\def\udtw{\smash{\rlap{\lower6pt\hbox{$\approx$}}}}

\def\tens#1{\ifmmode\mathchoice{\mbox{$\sf\displaystyle#1$}}
{\mbox{$\sf\textstyle#1$}}
{\mbox{$\sf\scriptstyle#1$}}
{\mbox{$\sf\scriptscriptstyle#1$}}\else
\hbox{$\sf\textstyle#1$}\fi}
\def\vec#1{\ifmmode\mathchoice{\mbox{\boldmath$\displaystyle#1$}}
{\mbox{\boldmath$\textstyle#1$}}
{\mbox{\boldmath$\scriptstyle#1$}}
{\mbox{\boldmath$\scriptscriptstyle#1$}}\else
\hbox{\boldmath$\textstyle#1$}\fi}
\def\bbbc{{\mathchoice {\setbox0=\hbox{$\displaystyle\rm C$}\hbox{\hbox
to0pt{\kern0.4\wd0\vrule height0.9\ht0\hss}\box0}}
{\setbox0=\hbox{$\textstyle\rm C$}\hbox{\hbox
to0pt{\kern0.4\wd0\vrule height0.9\ht0\hss}\box0}}
{\setbox0=\hbox{$\scriptstyle\rm C$}\hbox{\hbox
to0pt{\kern0.4\wd0\vrule height0.9\ht0\hss}\box0}}
{\setbox0=\hbox{$\scriptscriptstyle\rm C$}\hbox{\hbox
to0pt{\kern0.4\wd0\vrule height0.9\ht0\hss}\box0}}}}
\def\bbbq{{\mathchoice {\setbox0=\hbox{$\displaystyle\rm
Q$}\hbox{\raise
0.15\ht0\hbox to0pt{\kern0.4\wd0\vrule height0.8\ht0\hss}\box0}}
{\setbox0=\hbox{$\textstyle\rm Q$}\hbox{\raise
0.15\ht0\hbox to0pt{\kern0.4\wd0\vrule height0.8\ht0\hss}\box0}}
{\setbox0=\hbox{$\scriptstyle\rm Q$}\hbox{\raise
0.15\ht0\hbox to0pt{\kern0.4\wd0\vrule height0.7\ht0\hss}\box0}}
{\setbox0=\hbox{$\scriptscriptstyle\rm Q$}\hbox{\raise
0.15\ht0\hbox to0pt{\kern0.4\wd0\vrule height0.7\ht0\hss}\box0}}}}
\def\bbbt{{\mathchoice {\setbox0=\hbox{$\displaystyle\rm
T$}\hbox{\hbox to0pt{\kern0.3\wd0\vrule height0.9\ht0\hss}\box0}}
{\setbox0=\hbox{$\textstyle\rm T$}\hbox{\hbox
to0pt{\kern0.3\wd0\vrule height0.9\ht0\hss}\box0}}
{\setbox0=\hbox{$\scriptstyle\rm T$}\hbox{\hbox
to0pt{\kern0.3\wd0\vrule height0.9\ht0\hss}\box0}}
{\setbox0=\hbox{$\scriptscriptstyle\rm T$}\hbox{\hbox
to0pt{\kern0.3\wd0\vrule height0.9\ht0\hss}\box0}}}}
\def\bbbs{{\mathchoice
{\setbox0=\hbox{$\displaystyle     \rm S$}\hbox{\raise0.5\ht0\hbox
to0pt{\kern0.35\wd0\vrule height0.45\ht0\hss}\hbox
to0pt{\kern0.55\wd0\vrule height0.5\ht0\hss}\box0}}
{\setbox0=\hbox{$\textstyle        \rm S$}\hbox{\raise0.5\ht0\hbox
to0pt{\kern0.35\wd0\vrule height0.45\ht0\hss}\hbox
to0pt{\kern0.55\wd0\vrule height0.5\ht0\hss}\box0}}
{\setbox0=\hbox{$\scriptstyle      \rm S$}\hbox{\raise0.5\ht0\hbox
to0pt{\kern0.35\wd0\vrule height0.45\ht0\hss}\raise0.05\ht0\hbox
to0pt{\kern0.5\wd0\vrule height0.45\ht0\hss}\box0}}
{\setbox0=\hbox{$\scriptscriptstyle\rm S$}\hbox{\raise0.5\ht0\hbox
to0pt{\kern0.4\wd0\vrule height0.45\ht0\hss}\raise0.05\ht0\hbox
to0pt{\kern0.55\wd0\vrule height0.45\ht0\hss}\box0}}}}
\def\bbbz{{\mathchoice {\hbox{$\sf\textstyle Z\kern-0.4em Z$}}
{\hbox{$\sf\textstyle Z\kern-0.4em Z$}}
{\hbox{$\sf\scriptstyle Z\kern-0.3em Z$}}
{\hbox{$\sf\scriptscriptstyle Z\kern-0.2em Z$}}}}
\def\diameter{{\ifmmode\mathchoice
{\ooalign{\hfil\hbox{$\displaystyle/$}\hfil\crcr
{\hbox{$\displaystyle\mathchar"20D$}}}}
{\ooalign{\hfil\hbox{$\textstyle/$}\hfil\crcr
{\hbox{$\textstyle\mathchar"20D$}}}}
{\ooalign{\hfil\hbox{$\scriptstyle/$}\hfil\crcr
{\hbox{$\scriptstyle\mathchar"20D$}}}}
{\ooalign{\hfil\hbox{$\scriptscriptstyle/$}\hfil\crcr
{\hbox{$\scriptscriptstyle\mathchar"20D$}}}}
\else{\ooalign{\hfil/\hfil\crcr\mathhexbox20D}}%
\fi}}
\def\sq{\ifmmode\squareforqed\else{\unskip\nobreak\hfil
\penalty50\hskip1em\null\nobreak\hfil\squareforqed
\parfillskip=0pt\finalhyphendemerits=0\endgraf}\fi}
\def\squareforqed{\hbox{\rlap{$\sqcap$}$\sqcup$}}
\input{isolatin.sty}

\newcommand{\mciv}[1]{\multicolumn{4}{c}{#1}}
\newcommand{\BPRP}{\mbox{$G_{\rm BP}-G_{\rm RP}$}} 
\newcommand{\GG}{\mbox{$G$}}
\newcommand{\GBP}{\mbox{$G_{\rm BP}$}} 
\newcommand{\GRP}{\mbox{$G_{\rm RP}$}} 
\newcommand{\GBPc}{\mbox{$G_{\rm BP}^\prime$}} 
\newcommand{\GRPc}{\mbox{$G_{\rm RP}^\prime$}} 

\newcommand{\Teff}{\mbox{$T_{\rm eff}$}}

\newcommand{\chired}{\mbox{$\chi_{\rm red}$}}

\begin{document}
\pagenumbering{arabic}
\pagestyle{myheadings}
\thispagestyle{empty}
{\flushleft\includegraphics[width=\textwidth,bb=58 650 590 680]{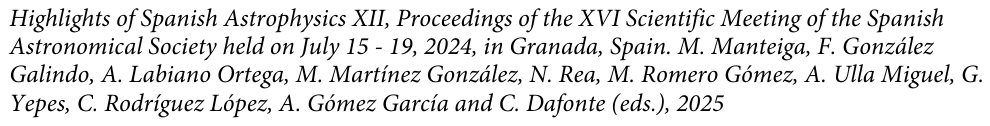}}
\vspace*{-1.0cm}


\begin{flushleft}
{\bf {\LARGE
%
Differences between Gaia DR2 and Gaia EDR3 photometry: 
demonstration, consequences, and applications
%
}\\
\vspace*{1cm}
%
J. Maíz Apellániz$^1$
and
M. Weiler$^2$
%
}\\
\vspace*{0.5cm}
%
$^1$ 
Centro de Astrobiolog{\'\i}a, CSIC-INTA, Spain\\
$^2$
Universitat de Barcelona, Spain\\
%
\end{flushleft}
%
\markboth{
Gaia DR2 and Gaia EDR3 photometry
}{ 
%
Ma{\'\i}z Apell\'aniz and Weiler    
%
}
\thispagestyle{empty}
\vspace*{0.4cm}
\begin{minipage}[l]{0.09\textwidth}
\ 
\end{minipage}
\begin{minipage}[r]{0.9\textwidth}
\vspace{1cm}
\section*{Abstract}{\small
%
We produce a new photometric calibration for the combined six-filter system formed by \textit{Gaia}~DR2~+~EDR3 \GG+\GBP+\GRP\ using an improved STIS/HST 
spectrophotometric library with very red stars. The comparison between observed and synthetic photometry yields residual dispersions of just
3.4-8.7~mmag, resulting in the most accurate and precise whole-sky large-dynamic-range optical photometric system ever obtained. We include some tests and 
applications.
%
\normalsize}
\end{minipage}
%
%
%

\section{Motivation and summary}

$\,\!$\indent \textit{Gaia} has provided us with very stable, whole-sky, high-dynamic-range, optical photometry for over $10^9$ stars but its full potential
can only be achieved if we produce a calibration as free as possible of systematic effects and with uncertainties that allow for a true comparison between
the observed magnitudes and synthetic photometry derived from SED models. In \citet{MaizWeil18} we obtained a photometric calibration for 
\textit{Gaia}~DR2 \GG+\GBP+\GRP\ that was proven to be the most accurate one at that point. We have repeated the procedure for \textit{Gaia}~EDR3 with an 
improved spectrophotometric library and also reanalyzed the original \textit{Gaia}~DR2 system. The result is the most stable, accurate, and precise 6-filter optical 
photometry system ever achieved. There are significant differences between the DR2 and EDR3 photometric systems that can be exploited for scientific 
purposes.


\begin{table}
\caption{Calibration summary. Specific passbands are applicable only to the magnitude range and common ones are magnitude independent. Magnitude corrections
can be global or applied only to saturated (very bright) stars. $Z_{\rm Vega}$ is the zero point for the passband in the Vega system. $\sigma_0$ is the 
minimum photometric uncertainty or value to add in quadrature to the catalog uncertainty when comparing observed magnitudes with synthetic photometry. }
\centerline{
\addtolength{\tabcolsep}{-2pt}
\begin{tabular}{ccccccccccc}
\\
\hline
Band & Range         & \mciv{DR2}                                          & & \mciv{EDR3}                                        \\
\cline{3-6} \cline{8-11}
     &               & passband & correction & $Z_{\rm Vega}$ & $\sigma_0$ & & passband & correction &$Z_{\rm Vega}$ & $\sigma_0$ \\
     & (mag)         &          &            & (mag)          & (mmag)     & &          &            & (mag)         & (mmag)     \\
\hline
\GG  & $\GG < 13.00$ & specific & yes        & $+$0.036       & 6.5        & & specific & yes        & $+$0.031      & 4.8        \\
     & $\GG > 13.00$ & specific & yes        & $+$0.041       & 5.1        & & specific & yes        & $+$0.033      & 3.9        \\
\hline
\GBP & $\GG < 10.87$ & specific & no         & $+$0.035       & 4.5        & & common   & saturation & $+$0.020      & 3.8        \\
     & $\GG > 10.87$ & specific & no         & $+$0.022       & 5.3        & & common   & no         & $+$0.020      & 4.0        \\
\hline
\GRP & $\GG < 10.87$ & specific & no         & $+$0.030       & 8.7        & & common   & saturation & $+$0.022      & 5.7        \\
     & $\GG > 10.87$ & specific & no         & $+$0.023       & 5.1        & & common   & no         & $+$0.022      & 3.4        \\
\hline
\end{tabular}
\addtolength{\tabcolsep}{2pt}
}
\label{maintable}                  
\end{table}


\section{Method}      

\begin{figure}
 \centerline{\includegraphics[width=\textwidth]{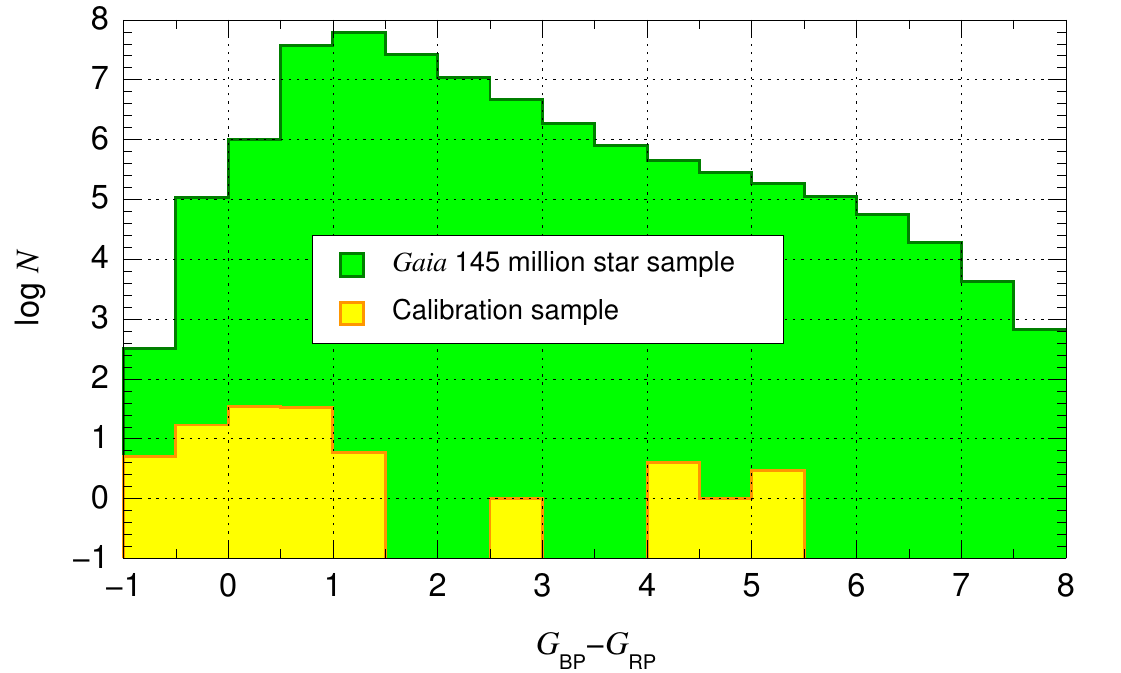}}
 \caption{$\GBP-\GRP$ histograms for the \textit{Gaia} 145 million star sample of \citet{Maizetal23} and for the calibration sample in this contribution.}
 \label{GBPmGRP_histo.pdf}   
\end{figure}

$\,\!$\indent In \citet{MaizWeil18} we compiled a spectrophotometric stellar library from high-quality STIS G430L+G750L data for 122 stars with a broad range
of colors and used the \citet{Weiletal18} technique to produce a full photometric calibration for \textit{Gaia} DR2 \GG+\GBP+\GRP. In this
contribution we have modified that list, eliminating some stars that were shown to have problems and, more importantly, adding stars from a dedicated
STIS program (GO 15\,816) to include additional very red stars: the previous list only had 3 stars with $2.9 < \GBP-\GRP < 4.8$ and none redder than that
while the new one has 5 stars with $2.9 < \GBP-\GRP < 4.8$ and 4 redder than that. Including a significant number of very red stars is crucial in 
determining the red end of the \GG\ and \GRP\ passbands and most spectrophotometric libraries do not have them. As the \textit{Gaia} population includes a significant
fraction of very red stars (Fig.~\ref{GBPmGRP_histo.pdf}), most of them extinguished red giants and red clump stars, a calibration without such very red stars is 
likely to provide incorrect information about them. With that improved list in hand, we have recomputed the \textit{Gaia}~DR2 and EDR3 passbands for \GG+\GBP+\GRP\ 
and produced a new photometric calibration for the 6-filter system. This contribution contains our preliminary results and the final results will appear in Weiler 
et al. (in preparation).


\section{Magnitude ranges and corrections}      

\begin{figure}
 \centerline{\includegraphics[width=\textwidth]{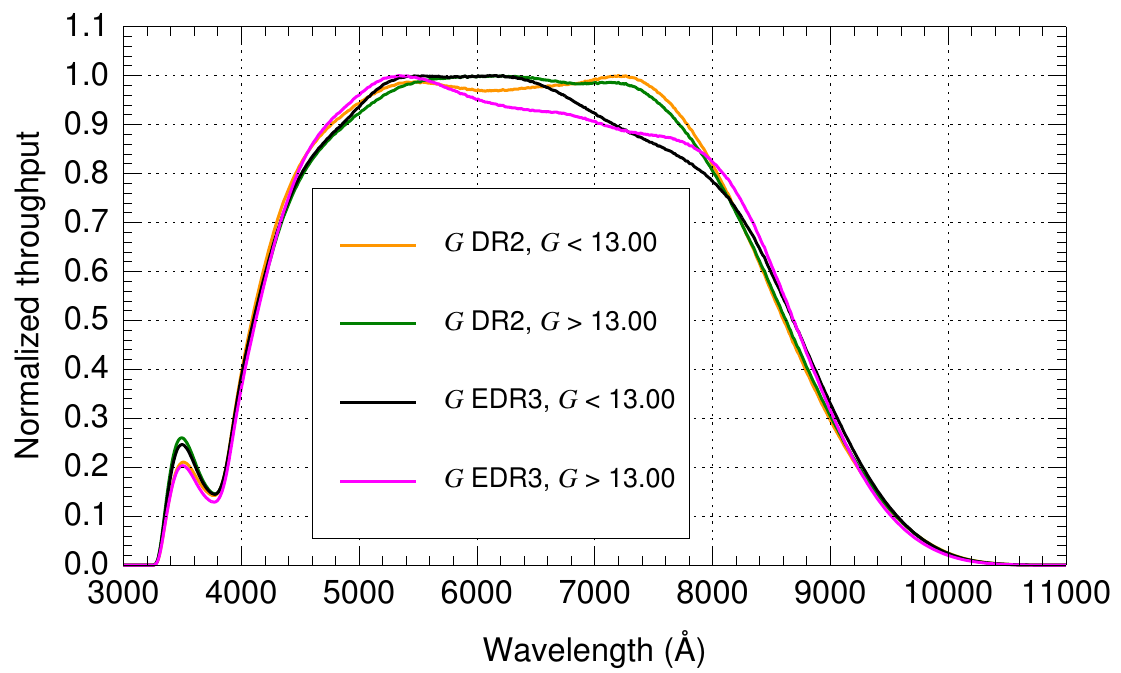}}
 \caption{\GG\ passbands.}
 \vspace{-5mm}
 \label{G}   
\end{figure}

\begin{figure}
 \centerline{\includegraphics[width=\textwidth]{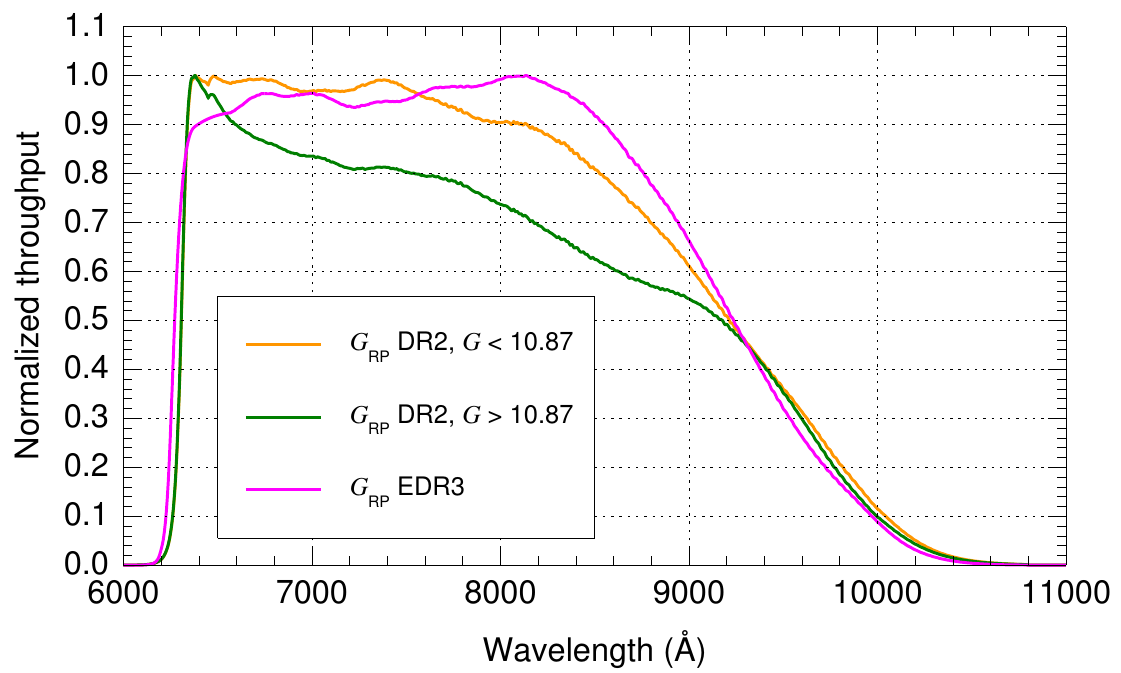}}
 \caption{\GRP\ passbands.}
 \vspace{-5mm}
 \label{GRP}   
\end{figure}

$\,\!$\indent In \citet{MaizWeil18} we computed \textbf{two \GBP\ DR2 bandpasses divided by \GG\ greater or smaller than 10.87~mag}, as we found out that the
sensitivity was different in those two magnitude ranges. In our new analysis of DR2 we confirm this finding and \textbf{we add an equivalent division at the
same \GG\ magnitude for \GRP} (Table~\ref{maintable} and Figs.~\ref{GBP}~and~\ref{GRP}). Furthermore, we have discovered that \textbf{the fit for the \GG\ band in 
both DR2 and EDR3 can be improved by adding a similar division at \GG~=~13.00~ma}g (Table~\ref{maintable}), though in this case the differences between
the bright and faint bandpasses are smaller than for \GBP\ or \GRP\ (Fig.~\ref{G}). Therefore, a total of ten different bandpasses are needed to characterize the
six-filter photometric system. The divisions in magnitude are likely caused by the use of TDI gates for the
processing of \textit{Gaia} CCD data \citep{Prusetal16}, so the origin is electronic and not optical. 

In \citet{MaizWeil18} we also included \textbf{a correction $\delta G_2$ for the \GG\ DR2 band} that led to corrected magnitudes 
$G^\prime_2 = G_2 + \delta G_2$ (see also \citealt{Arenetal18,Weil18,CasaVand18}). We have recomputed the correction here and found a similar result 
(Fig.~\ref{Gcor}). Furthermore, we find that an equivalent correction is needed for the \GG\ EDR3 band (Fig.~\ref{Gcor}). For \GBP\ and \GRP\ we find that
the EDR3 saturation corrections of \citet{Rieletal21} (the results from their application may be called \GBPc\ and \GRPc, respectively) are sufficient.


\section{Passband definitions and evolution}      

\begin{figure}
 \centerline{\includegraphics[width=\textwidth]{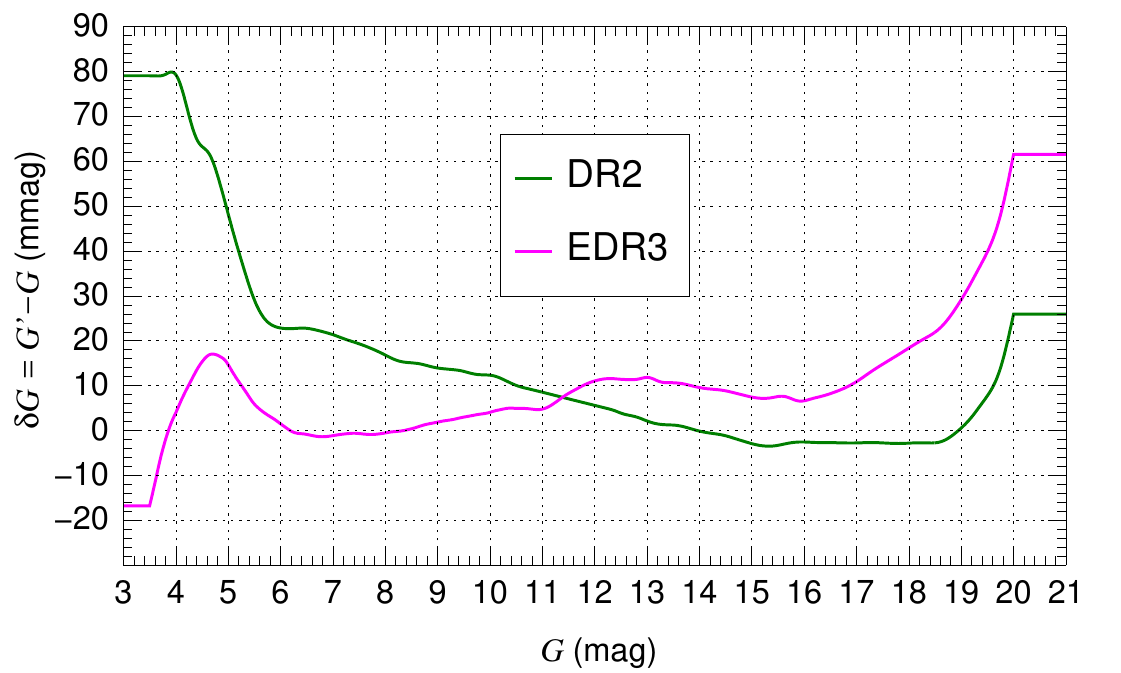}}
 \vspace{-5mm}
 \caption{\GG\ corrections as a function of magnitude.}
 \label{Gcor}   
\end{figure}

$\,\!$\indent In \citet{Maiz17a} we discovered that the \GG\ DR1 passband was significantly redder than the nominal one from \citet{Jordetal10}. This was
likely caused by water freezing in some optical elements at the beginning of the mission \citep{Prusetal16}. In \citet{MaizWeil18} we detected that the
\GG\ DR2 passband had become bluer, as the DR2 magnitudes were computed using a longer time baseline, and the effect of ice deposition had become less
significant on average. Here we confirm the trend in time, as \textbf{for EDR3 the two \GG\ passbands are even bluer than their DR2 equivalents 
(Fig.~\ref{Gcor})}.

\begin{figure}
 \centerline{\includegraphics[width=\textwidth]{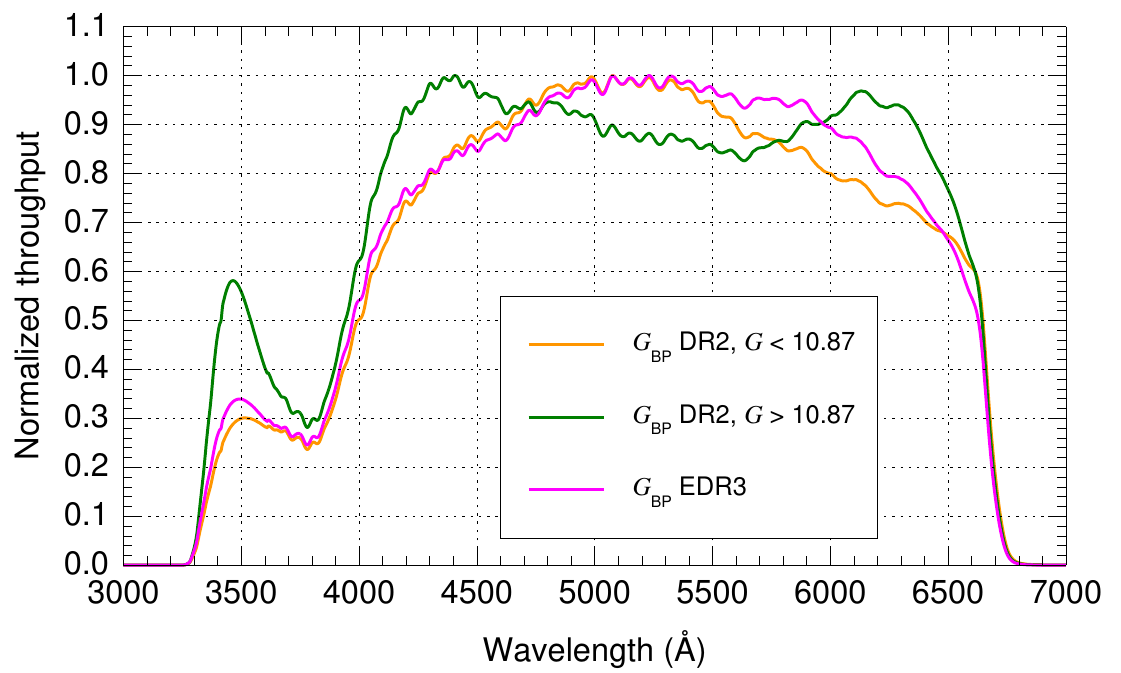}}
 \caption{\GBP\ passbands.}
 \vspace{-5mm}
 \label{GBP}   
\end{figure}

In \citet{MaizWeil18} we found out that the bright ($G < 10.87$~mag) \GBP\ DR2 passband was different from the faint ($G > 10.87$~mag) one, most
significantly in its lower sensitivity to the left of the Balmer jump. Here we confirm that and we determine that the \GBP\ EDR3 passband is relatively 
similar to the bright DR2 one (and hence, significantly different from the faint one, Fig.~\ref{GBP}). \textbf{The difference between the DR2 and EDR3 
 \GBP\ bandpasses for faint stars is important, given its scientific applications} (Figs.~\ref{GBPmGRP_deltaGBP}~and~\ref{GBPmGRP_deltaGBP2}, see below).

The \citet{MaizWeil18} analysis did not include two passbands for \GRP\ in DR2 but the addition of very red stars to our new sample allows us
to establish that they are indeed different, with the passband for faint ($G > 10.87$~mag) stars being significantly bluer than the one for bright
stars. Furthermore, as we saw for \GBP, the \GRP\ EDR3 bandpass is relatively similar to the DR2 one for bright stars and significantly different from the
DR2 one for faint stars. \textbf{The difference between the DR2 and EDR3 \GRP\ bandpasses for faint stars can also be exploited}, as it is sensitive in
different degrees to broad-band colors in the 7000-9000~\AA\ region and to H$\alpha$.


\section{Accuracy and precision}

$\,\!$\indent In the two previous sections we have described the corrections that have to be applied to the observed \GG\ photometry, the magnitude ranges in
which each filter has to be divided, and the respective passbands. Two final steps are required for a meaningful comparison between the observed photometry and the 
synthetic one obtained from SED models.

First, a Vega zero point, $Z_{\rm Vega}$, has to be determined for each passband (Table~\ref{maintable}) in order for accurate synthetic magnitudes to be obtained through:

\begin{equation}
m_{\rm syn} = -2.5 \log_{10} 
  \left(\frac{\int P(\lambda)\,F_\lambda(\lambda)             \,\lambda\,d\lambda}
             {\int P(\lambda)\,F_{\lambda,{\rm Vega}}(\lambda)\,\lambda\,d\lambda}\right) 
  + {\rm Z}_{\rm Vega},
\label{msyn}
\end{equation}

\noindent see \citet{Maiz24} for definitions and details. In particular, we adopt the Vega SED \texttt{alpha\_lyr\_stis\_010.fits} from 
\href{https://www.stsci.edu/hst/instrumentation/reference-data-for-calibration-and-tools/astronomical-catalogs/calspec}{CALSPEC} \citep{Bohl14}.

Second, the overall precision of the calibration has to be established by determining the minimum photometric uncertainty $\sigma_0$ to be combined with the observed
photometric uncertainties $\sigma$ to generate the corrected uncertainties $\sigma^\prime = \sqrt{\sigma^2 + \sigma_0^2}$. The corrected uncertainties are the relevant ones
for a comparison between observed and synthetic magnitudes and include the uncertainties in our knowledge of the passbands and other peculiarities of the photometric system.
They are calculated by comparing the (assumed exact) synthetic magnitudes from HST spectrophotometry with the observed magnitudes divided by the corrected uncertainties and forcing 
the resulting distribution to have a standard deviation of one (the mean is already forced to be zero by the calibration itself). 

The values of $\sigma_0$ are very low (Table~\ref{maintable}), ranging from 4.5~mmag to 8.7~mmag for DR2 and from 3.4~mmag to 5.7~mmag for EDR3. They are significantly lower than 
their equivalents for other photometric systems \citep{Maiz06a}, thus \textbf{establishing \textit{Gaia} as the the most precise whole-sky large-dynamic-range optical photometric 
system ever obtained.} Furthermore, the improvement in the calibration from DR2 to EDR3 is noticeable in the reduction of the $\sigma_0$ values.


\section{Testing and applications}

$\,\!$\indent We tested our passbands with a large sample of OB stars from the ALS project (\citealt{Pantetal21}; 2024 in preparation) and synthetic photometry 
computed with the SED grid of \citet{Maiz13a}. The use of OB stars allows us to select targets with similar intrinsic SEDs (with the main difference being at the left 
of the Balmer jump) modified by different levels of extinction to generate observed SEDs of different colors.


\begin{figure}
 \centerline{\includegraphics[width=\textwidth]{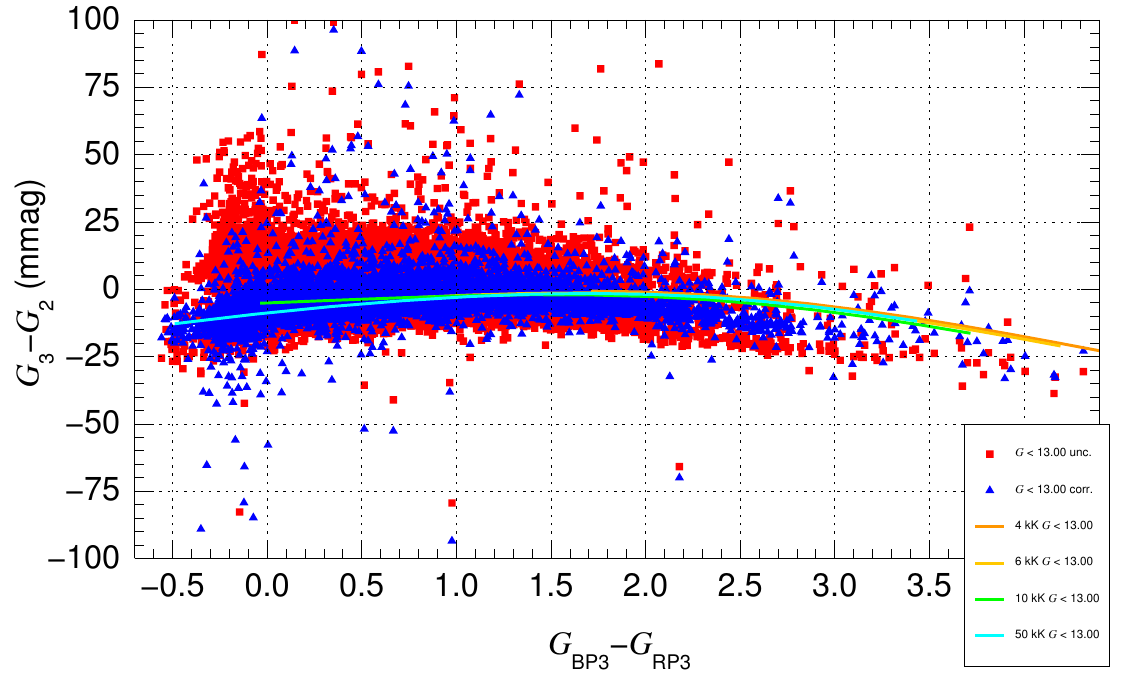}}
 \caption{$\Delta G$ between EDR3 and DR2 as a function of $\GBP-\GRP$ for the ALS+ sample with $G< 13.00$ using uncorrected (red) and corrected (blue)
          magnitudes. The lines show the extinction tracks \citep{Maizetal14a} for MS stars of different \Teff.}
 \label{GBPmGRP_deltaG}   
\end{figure}

First, we plot the difference between the EDR3 and DR2 \GG\ magnitudes as a function of \BPRP\ using both uncorrected and corrected magnitudes (Fig.~\ref{GBPmGRP_deltaG}).
The corrections significantly reduce the dispersion, confirming their validity. In addition, the distribution follows the function expected from the synthetic photometry of 
extinguished OB stars, confirming the validity of the passbands.

\begin{figure}
 \centerline{\includegraphics[width=\textwidth]{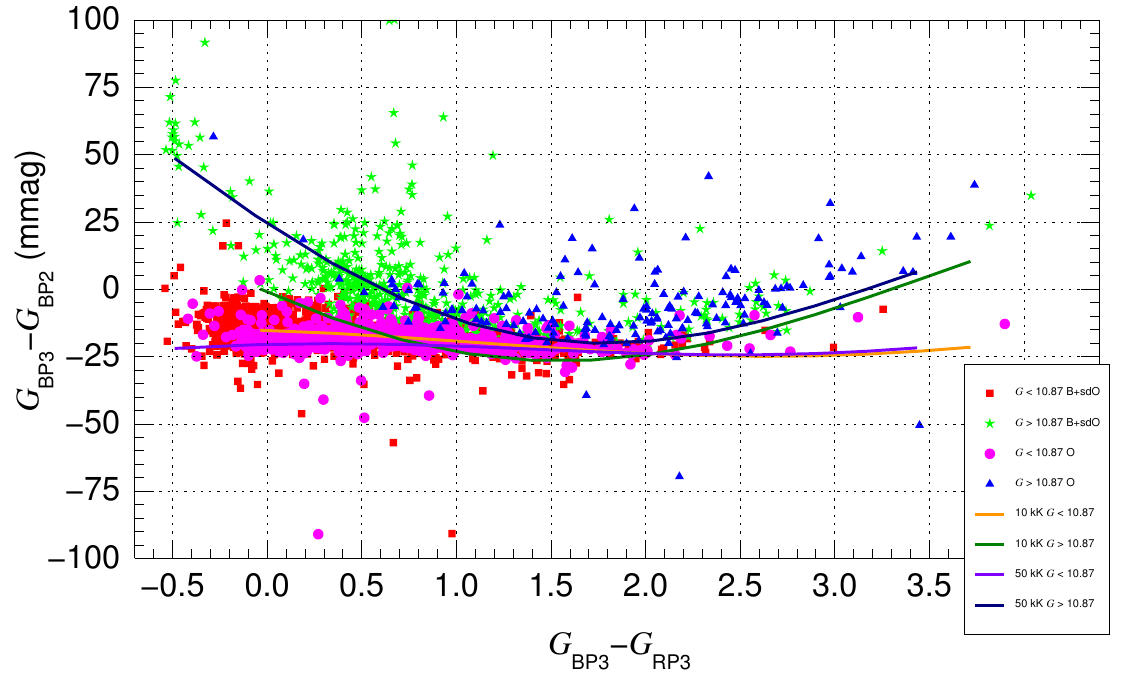}}
 \caption{$\Delta \GBP$ between EDR3 and DR2 as a function of $\GBP-\GRP$ for the ALS+ early-type samples 
          The lines show the extinction tracks for MS stars of different \Teff.}
 \label{GBPmGRP_deltaGBP}   
\end{figure}

In Fig.~\ref{GBPmGRP_deltaGBP} we plot the difference between the EDR3 and DR2 \GBP\ magnitudes as a function of \BPRP\ dividing the sample in four according to two criteria: \GG\
brighter or fainter than $\GG = 10.87$~mag and spectral clssification as O or B (with the latter including a small number of sdO stars). In that plot:

\begin{itemize} 
 \item Bright stars ($\GG< 10.87$~mag) follow a nearly constant trend in color, as predicted by the synthetic photometry of stars with \Teff\ in the 10-50~kK range. The explanation
       is that the \GBP\ bright DR2 and EDR3 passbands are very similar (Fig.~\ref{GBP}), with an offset created by the different zero points (Table~\ref{maintable}).
 \item Faint stars ($\GG> 10.87$~mag), on the other hand, have a significant dependence with color, also in accordance with the synthetic photometry. This is explained by the
       difference between the \GBP\ faint DR2 and EDR3 passbands (Fig.~\ref{GBP}): the former is more sensitive for $\lambda < 4700$~\AA, especially for the region around 3500~\AA\
       that lies to the left of the Balmer jump. For hot stars with low extinction, a significant fraction of the flux comes from that region, causing the DR2 magnitude to be brighter
       than the EDR3 one in Fig.~\ref{GBPmGRP_deltaGBP}. As OB stars become more extinguished, the contribution from that wavelength region becomes less important and the two
       magnitudes become similar.
 \item Another effect for the faint range is that for highly extinguished OB stars ($\BPRP > 2.0$~mag) the above trend is reversed and DR2 \GBP\ magnitudes become once again brighter
       than EDR3 ones. This also has an explanation in Fig.~\ref{GBP}. At the red end of the filter ($\lambda > 6000$~\AA), the faint DR2 passband becomes more sensitive than the EDR3 
       one. When OB stars are highly extinguished, a significant fraction of the flux originates in that part of the passband, causing the effect.
\end{itemize}

\begin{figure}
 \centerline{\includegraphics[width=\textwidth]{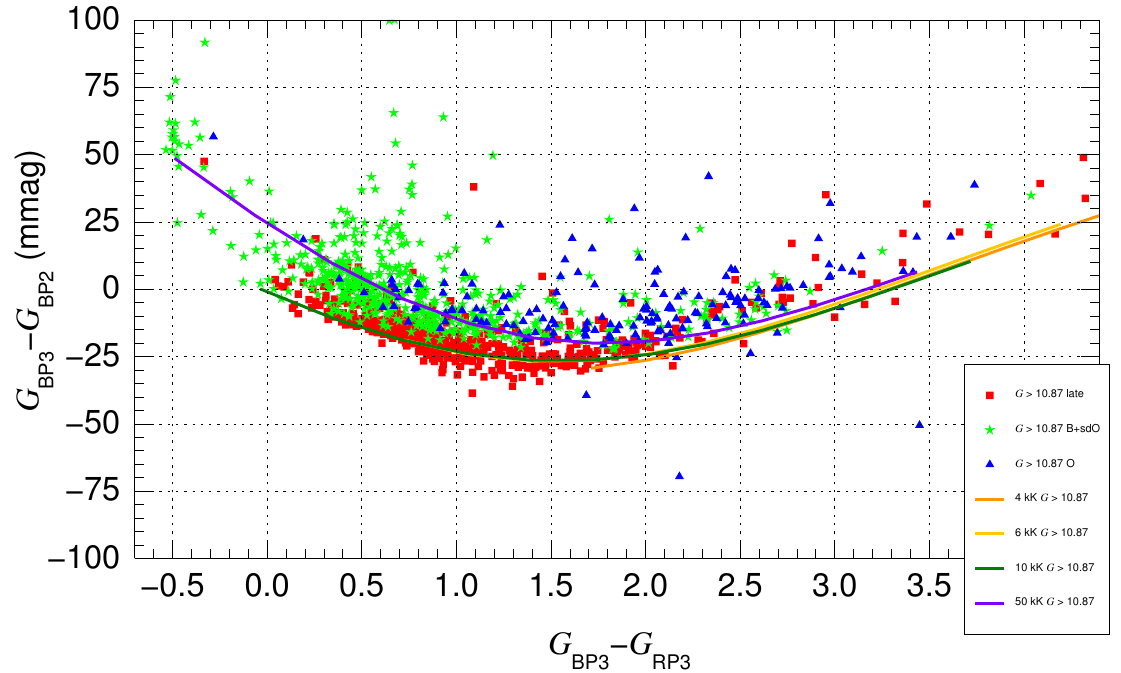}}
 \caption{$\Delta \GBP$ between EDR3 and DR2 as a function of $\GBP-\GRP$ for the ALS+ sample with $G> 10.87$.
          The lines show the extinction tracks for MS stars of different \Teff.}
 \label{GBPmGRP_deltaGBP2}   
\end{figure}

In Fig.~\ref{GBPmGRP_deltaGBP2} we plot only the faint OB stars in Fig.~\ref{GBPmGRP_deltaGBP} and we also add a sample of later-type objects dominated by A stars. There we see a
significant \Teff\ trend, with hotter stars being brighter in DR2 than in EDR3 for a given \BPRP\ color. The trend is more marked at low extinction, as the effect of the flux to the
left of the Balmer jump is stronger there (see above) and eventually disappears at very high extinction (note that there are more low-extinction B stars than the equivalent of O type,
a well known effect among Galactic bright stars). Therefore, \textbf{it is possible to use the \textit{Gaia} six-filter system to estimate \Teff\ among OBA stars} in a manner analogous
to classical Johnson $U-B$ versus $B-V$ diagrams.

A test of the quality of the calibration can be done using the SED-fitting code CHORIZOS \citep{Maiz04c}, which has been adapted to include it. 
We have applied it to several OB stars of different degrees of extinction in \citet{Maizetal24a} and in poster P252 in these proceedings \citep{MaizNegu25} 
in combination with 2MASS $JHK$ photometry to produce a nine-degree filter system and
simultaneously derive the extinction parameters and the luminosity. The resulting fits are excellent, with \chired\ values close to 1.0, something that can only happen if the
calibration, intrinsic SEDs, and extinction laws are all simultaneously accurate \citep{Maiz24}.


\section{Data availability}

$\,\!$\indent We have integrated our analysis into an IDL package called Gaiasoft that downloads \textit{Gaia} data from VizieR, introduces the 
necessary corrections, plots the passbands, and computes the synthetic photometry. It will be released when Weiler et al. is published. In the meantime, 
contact the main author at \href{mailto:jmaiz@cab.inta-csic.es}{jmaiz@cab.inta-csic.es} for conditions on its use.


\bibliographystyle{aa}
\bibliography{general}

\end{document}